\documentclass{article}
\usepackage[T1]{fontenc}
\usepackage[utf8]{inputenc}
\usepackage{ismir}
\usepackage{amsmath,cite,url}
\usepackage{graphicx}
\usepackage{color}
\usepackage{amsfonts}
\usepackage{enumitem}
\usepackage{algorithm}
\usepackage{algorithmic}
\usepackage{amssymb}
\usepackage{booktabs}
\usepackage{pifont}

\usepackage{multirow}

\title{Fx-Encoder++: Extracting Instrument-Wise Audio Effects Representations from Mixtures}

\multauthor
  {*Yen-Tung Yeh$^1$\thanks{*Work done during an internship at Sony AI} \hspace{1cm} Junghyun Koo$^2$ \hspace{1cm} Marco A. Martínez-Ramírez$^2$ }
  {{\bf Wei-Hsiang Liao$^2$ \hspace{1cm} Yi-Hsuan Yang$^1$ \hspace{1cm} Yuki Mitsufuji$^{2,3}$}\\
  $^1$ National Taiwan University, Taipei, Taiwan, \; \\
  $^2$ Sony AI, Tokyo, Japan, \; 
  $^3$ Sony Group Corporation, Tokyo, Japan \; \\
  {\tt\small f12942179@ntu.edu.tw}
  }

\def\authorname{Y-T. Yeh, J. Koo, M. Martínez-Ramírez, W-H. Liao, Y-H Yang, and Y. Mitsufuji}

\usepackage[bookmarks=false,pdfauthor={\authorname},pdfsubject={\pdfsubject},hidelinks]{hyperref}

\begin{document}

\maketitle

\begin{abstract}
General-purpose audio representations have proven effective across diverse music information retrieval applications, yet their utility in intelligent music production remains limited by insufficient understanding of audio effects (Fx). Although previous approaches have emphasized audio effects analysis at the mixture level, this focus falls short for tasks demanding instrument-wise audio effects understanding, such as automatic mixing. In this work, we present \textbf{Fx-Encoder++}, a novel model designed to extract instrument-wise audio effects representations from music mixtures. Our approach leverages a contrastive learning framework and introduces an ``extractor'' mechanism that, when provided with instrument queries (audio or text), transforms mixture-level audio effects embeddings into instrument-wise audio effects embeddings. We evaluated our model across retrieval and audio effects parameter matching tasks, testing its performance across a diverse range of instruments. The results demonstrate that Fx-Encoder++ outperforms previous approaches at mixture level and show a novel ability to extract effects representation instrument-wise, addressing a critical capability gap in intelligent music production systems.

\end{abstract}

\section{Introduction}\label{sec:introduction}

Recent advances in deep learning have enabled significant progress in general-purpose audio representations \cite{hershey2017cnn, kong2020panns, wu2023large, defossez2022high, chen2022beats}, which have proven effective across diverse applications. However, they inadequately capture the nuanced characteristics of audio effects processing \cite{steinmetz2024st, hawley2023leveraging, chu2025text2fx}, since they prioritize semantic content recognition over subtle audio effects or timbral transformations. This shortcoming particularly affects specialized applications that require a precise understanding of audio effects configuration and their perceptual impact across different musical contexts.

A key domain affected by this representation gap is intelligent music production, the emerging field of mixing and mastering automation aiming to develop AI systems capable of professional-quality audio engineering \cite{de2019intelligent}. Applications in this domain, including automatic mixing \cite{steinmetz2021automatic, martinez2021deep}, audio effects style transfer \cite{steinmetz2022style, steinmetz2024st, koo2022end}, and mixing style transfer \cite{vanka2024diff, koo2023music}, all require specialized representations that model effects at two distinct levels: how they shape the overall sound of a complete mixture (``mixture level'') and how they transform individual instruments within that mixture (``instrument level'').

Previous approaches to audio effects representation can be categorized into two groups: those analyzing effects at the entire musical mixes \cite{koo2023music, steinmetz2022style} and those examining effects on isolated single instruments \cite{chen2024towards, wright2025open}. Although both approaches operate at what we consider the ``mixture level'', they fail to extract instrument-specific effect characteristics from complex mixtures. 
To our knowledge, FX-Encoder \cite{koo2023music} is the only existing work that encodes audio effects information from individual instrument stems. 
However, they focus only on modeling the aggregate result, rather than identifying how effects have been applied to each instrument. This limitation restricts applications such as automatic mixing, where a precise understanding of the instrument-wise processing is essential.

Extracting instrument-specific audio effects representations from mixtures presents two distinct challenges. First, a straightforward approach might involve applying source separation algorithms \cite{kong2023universal, defossez2019music} to isolate individual instruments before analyzing their effects. This method is limited to separation artifacts, such as missing high frequencies, transient smearing \cite{schaffer2022music}, and imperfect signal reconstruction. Those artifacts will distort effect characteristics, making the separate-then-analyze pipeline unreliable. Second, to obtain accurate effect parameters, we require both the processed track (wet audio) and its unprocessed counterpart (dry audio) for reverse engineering \cite{colonel2021reverse, lee2024searching}. This necessitates not only perfect source separation to extract the processed track from the mixture but also access to the original unprocessed audio, which is rarely available. 


To address these limitations, we propose \textbf{Fx-Encoder++}, a novel model designed to extract instrument-specific audio effects representations from complete music mixtures. Our model employs a contrastive learning framework based on SimCLR \cite{chen2020simple}, which learns representations by maximizing agreement between different augmented views of the same data. Following FX-Encoder \cite{koo2023music}, we implement several crucial design elements to ensure effective learning for audio effects including: i) Fx-Normalization \cite{martinez2022automatic} to normalize inherent effects in source audio; ii) consistent instrumentation composition, to ensure each mixture in the entire batch is constructed with the same instrument combination, thereby isolating effect-related features; and iii) systematic audio effects manipulation procedures. The core of our contribution is an ``extractor'' mechanism that transforms mixture-level embeddings into instrument-specific embeddings when provided with instrument queries. This extractor leverages a pretrained CLAP encoder \cite{wu2023large} to support both audio and text queries.

In our experiments, we evaluate Fx-Encoder++ against both general-purpose audio representations (VGGish \cite{hershey2017cnn}, PANN \cite{kong2020panns}, and CLAP \cite{wu2023large}) and specialized audio effects encoders (FX-Encoder \cite{koo2023music} and AFx-Rep \cite{steinmetz2024st}). Building upon the evaluation framework of AFx-Rep \cite{steinmetz2024st}, we introduce a comprehensive benchmark for audio effects retrieval using real-world multitrack recordings from MUSDB \cite{musdb18-hq} and MedleyDB \cite{bittner2014medleydb}. 
We quantitatively assess performance across two dimensions: effects complexity (single vs. multiple cascaded effects) and instrument generalization. Results demonstrate that Fx-Encoder++ consistently outperforms existing methods in mixture-level effects extraction tasks. 
Furthermore, our approach enables instrument-specific effects extraction, addressing a gap in previous work where effects could only be analyzed at the mixture level.

\begin{figure*}[t]
\centering
\includegraphics[width=.99\textwidth, height=6.5cm, keepaspectratio]{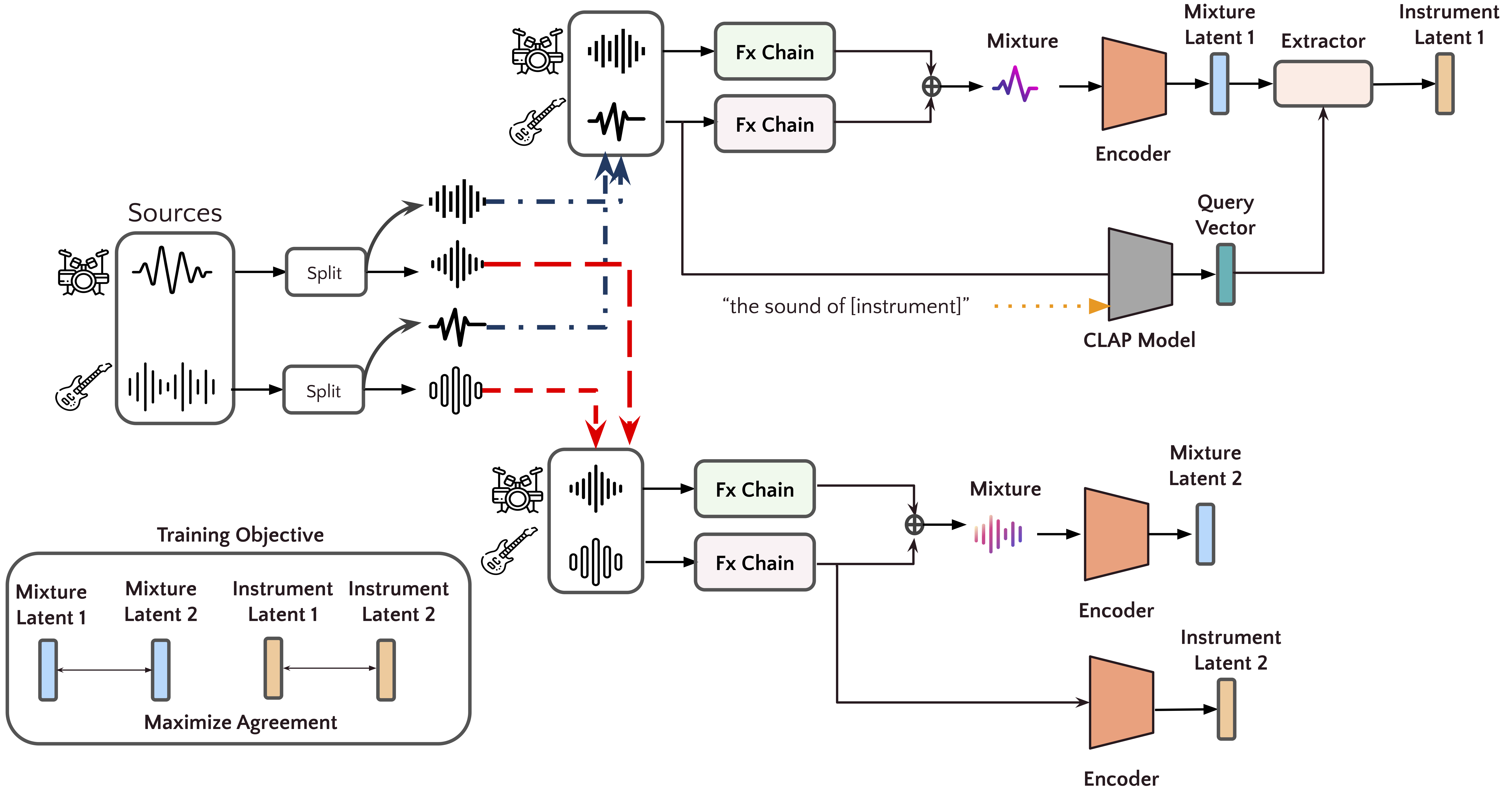}
\vspace{-2mm}
\caption{Our data preparation pipeline creates training pairs by splitting source tracks ($n=2$ in this example) into segments, applying consistent effect chains, and combining them into mixtures with identical effect configurations but distinct musical content. The architecture consists of an encoder that processes mixtures to produce mixture-level embeddings, and an extractor mechanism that transforms these embeddings into instrument-specific representations using CLAP-derived queries. During inference, we can optionally provide text query to extract the specific instruments.}
\label{fig:full_model_arc}
\end{figure*}

\section{Related works}

\subsection{General-purpose Audio Representation}

General-purpose audio representations have emerged to support various downstream tasks. VGGish \cite{hershey2017cnn} and PANN \cite{kong2020panns} employ CNNs trained on AudioSet \cite{gemmeke2017audio} for audio classification and pattern recognition. More recently, CLAP \cite{wu2023large} uses a transformer-based architecture with contrastive learning to align audio and text, enabling applications in text-to-music generation \cite{liu2024audioldm, copet2023simple} and audio separation \cite{liu2024separate, saijo2025leveraging}. Neural audio compression models \cite{defossez2022high, kumar2023high, pasini2024music2latent} represent another category, utilizing VAEs to reconstruct perceptual features while minimizing bitrate.
Despite their success across various tasks, these representations show limited sensitivity to audio effects transformations \cite{steinmetz2024st, hawley2023leveraging, chu2025text2fx}. This limitation stems from training objectives prioritizing semantic content recognition or cross-modal alignment rather than preserving the subtle timbral modifications introduced by audio effects, a capability essential for intelligent music production systems.

\subsection{Audio Effects Representation}
Audio effects representation research has evolved from implicit to explicit approaches. Early works incorporated effects awareness indirectly through applications like reverb parameter estimation \cite{koo2021reverb}, automatic mixing \cite{chen2022automatic}, differentiable effects \cite{lee2022differentiable}, and neural style transfer \cite{steinmetz2022style, vanka2024diff}.
Dedicated representations emerged with FX-Encoder \cite{koo2023music}, which used contrastive learning to disentangle effects characteristics from content, enabling mixing style transfer. Similarly, Tone Embedding \cite{chen2024towards} 
 and  OpenAMP \cite{wright2025open} focused specifically on guitar tones but was limited to isolated guitar recordings. Recently, AFx-Rep \cite{steinmetz2024st}, a classification-based model designed for inference-time effects optimization. Its training objective is to classify which single effect is applied between two given audio clips and to further predict which preset that effect belongs to. Despite these advances, a limitation persists: existing models operate either at mixture level or on isolated instruments, but cannot extract instrument-wise effects information from complete mixes. Table~\ref{tab:fx_models_comparison} compares audio effects representation approaches. ``Extraction Level'' indicates the source from which models extract effects representations: ``Isolated Inst.'' for single-instrument capability, ``Mixture'' for complete music mixes. Fx-Encoder++ uniquely extracts at both mixture and instrument-wise levels. The ``Query'' column shows whether the model supports conditional extraction based on instrument queries, a distinctive feature of our approach.



\begin{table}[t]
\centering
\footnotesize
\renewcommand{\arraystretch}{1.1}
\setlength{\tabcolsep}{3pt}
\resizebox{1.0\linewidth}{!}
{
\begin{tabular}{lcccc}
\toprule
Model & Training & Audio & Extraction & Query \\
       & Method & Type & Level &  \\
\midrule
FX-Encoder \cite{koo2023music} & Contrastive & Mixture & Mixture & - \\
Tone Emb. \cite{chen2024towards} & Contrastive & Guitar & Isolated Inst. & - \\
OpenAMP \cite{wright2025open} & Contrastive & Guitar & Isolated Inst. & - \\
AFx-Rep \cite{steinmetz2024st} & Classification & Mixture & Mixture & - \\
Fx-Encoder++ (ours) & Contrastive & Mixture & Mixture \& Inst. &  \checkmark \\
\bottomrule
\end{tabular}
}
\caption{Comparison of audio effect representation models. Fx-Encoder++ uniquely extracts instrument-specific effects directly from mixtures, unlike previous approaches limited to whole-mix effects or isolated instruments.}
\label{tab:fx_models_comparison}
\end{table}

\section{Method}

Our goal is to develop an encoder, denoted as $\mathcal{E}(\mathbf{x, q})$, that encodes an audio effects embedding from music mixtures $\mathbf{x}$. When conditioned with an instrument query $\mathbf{q}$, the encoder extracts effects representations specific to that instrument within the mixture; without conditioning (i.e., $\mathbf{q} = \emptyset$), the encoder produces representations that characterize the complete mixture. 
We formulate a contrastive objective for learning mixture-level representations following FX-Encoder \cite{koo2023music}, then extend it with a mechanism to extract instrument-specific effects information directly from mixtures, as shown in Figure \ref{fig:full_model_arc}.

\subsection{Contrastive Objective for Audio Effects} \label{section:objective_for_fx}

Following FX-Encoder \cite{koo2023music}, we employed a SimCLR-based \cite{chen2020simple} contrastive objective to learn audio effects representations by maximizing the agreement between different audio contents processed with identical effects. The contrastive loss for a positive pair ($i$, $j$) can be formulated as:

\begin{equation}
\ell_{i,j}^{\text{mixture}} = -\log \frac{\exp(\text{sim}(z_i, z_j)/\tau)}{\sum_{k=1}^{2N} \mathcal{I}_{[k\neq i]} \exp(\text{sim}(z_i, z_k)/\tau)}
\label{eq:contrast_loss}
\end{equation}
where $\mathcal{I}_{[k\neq i]} \in \{0, 1\}$ is an indicator function evaluating to 1 if $k \neq i$, $\tau$ denotes a temperature parameter, and embeddings $z_i = \mathcal{E}(M_i)$ and $z_j = \mathcal{E}(M_j)$ are encoder outputs from mixtures $M_i$ and $M_j$. These mixtures are generated detailed as follows:

\begin{itemize}[leftmargin=*,itemsep=0pt,topsep=2pt]
    \item \emph{Fx-Normalization \cite{martinez2022automatic}}. 
    Directly applying identical audio effects to different audio clips is insufficient for our contrastive objective due to inherent effects already present in each recording. When applying effects $f$ to clips $x_a$ and $x_b$ with underlying effects $f_a$ and $f_b$, we obtain distinct transformations $f \circ f_a$ and $f \circ f_b$ despite using the same effects $f$, contradicting our goal of clustering samples with the same effects in embedding space.
    To overcome this limitation, we employed Fx-Normalization data preprocessing method \cite{martinez2022automatic}, the technique normalized audio characteristics by neutralizing existing effects, ensuring subsequently applied effects make equal contributions across all samples. 

    \item \emph{Consistent Instrument Composition}. We ensure all data samples within each training batch share identical instrument composition, preventing the model from taking shortcuts by distinguishing between instrument combinations (e.g., ``drums+bass'' versus ``guitar+vocals'') rather than effects characteristics. By controlling instrumentation across both positive and negative pairs, we force the model to focus on subtle effects features instead of more easily distinguishable instrument timbres. This is crucial for SimCLR, where numerous negative samples could otherwise lead to exploitation of simpler patterns based on instrument content \cite{robinson2021can}.

    \item \emph{Audio Effects Manipulation}. We create mixture pairs with identical effects processing but different musical content through a systematic approach: (1) generate $k$ distinct audio effects chains by randomly sampling configurations (order, number, types, and parameters), from our effects pool (equalizer, delay, distortion, stereo imager, compression, limiter, reverb, and gain); (2) randomly select $k$
    instruments from the normalized pool; (3) split each instrument track into two segments with different musical content; (4) apply the same effects chain $E_j$ to both segments of instrument $j$; (5) create two mixtures—$M_1$ and $M_2$—by combining all first and second segments respectively, with random loudness normalization ($-18$ to $-14$ dB LUFS) for individual track and consistent $-18$ dB LUFS for final mixtures. This creates positive pairs with different content but identical effects processing for contrastive learning. We also employ Fx probability scheduling \cite{koo2023music} to prevent the model from focusing only on easily distinguishable effects.
    
    \item \emph{Hand-Crafted Hard Negative Samples}. Hard negative samples are crucial for effective contrastive learning \cite{robinson2020contrastive}. We explicitly construct hard negatives by applying different effects chains to identical source material, creating samples with same musical content but different effects styles. We ensure variation through both structural differences (altering effect processor types, counts, and ordering) and independent parameter sampling. This approach forces the model to focus specifically on effects-based rather than content-based features.
    
\end{itemize}

\subsection{Learnable Extractor}

A critical design requirement is producing both mixture-level and instrument-specific effects representations with high fidelity. We incorporate instrument queries from a pretrained CLAP encoder \cite{wu2023large}, allowing audio queries during training and text queries during inference \cite{saijo2025leveraging} for intuitive control. While we could condition the encoder using methods like FiLM \cite{perez2018film}, this creates a challenge: the appropriate query vector for extracting the ``global'' mixture effects. A zero vector lacks semantic meaning, while using the mixture itself as a query would force the model to simultaneously analyze and extract from the same information stream. 
Our architecture therefore addresses dual requirements: extracting instrument-specific components while maintaining mixture-level representation.

Inspired from \cite{chen2024contrastive}, our \emph{extractor} mechanism resolves this by maintaining separate paths for mixture-level effects and instrument-specific extraction. The base encoder learns effects representations from mixtures, while the extractor transforms these based on instrument queries (audio or text). This two-stage design creates a principled interface between effects representation and instrument content representation, enabling targeted extraction of effects characteristics for specific instruments.

To facilitate the transformation of mixture effects embeddings into instrument-specific effects embeddings, we reformulate the standard contrastive loss. Rather than operating solely on mixture pairs, we develop an instrument-aware contrastive objective based on triplets (${Q, M_i, M_j}$), where $Q$ represents the instrument query, and $M_i$ and $M_j$ denote music mixtures with identical effects configurations. To make this approach compatible with mixture-level training, we construct an \emph{extractor} mechanism that extracts instrument-specific effect embeddings from mixture-level effects embeddings using instrument queries. Formally, we define the extractor as:
\begin{equation}
z_i^m = \texttt{extractor}(Q_i^m, \mathcal{E}(M_i))
\end{equation}
where $z_i^m$ represents the instrument-specific effects embedding, $Q_{i}^{m} \in \mathbb{R}^D$ is the query vector for the $m$-th instrument in mixture $M_i$ (obtained from a pretrained CLAP encoder), and $\mathcal{E}(M_i)$ is the mixture-level audio effects embedding.  The extractor is implemented as a multi-layer perceptron that learns to selectively attend to effects-related features associated with the queried instrument. The instrument-aware contrastive loss is then formulated as:
\begin{equation}
\ell_{i,j}^{\text{inst}} = -\log \frac{\exp(\text{sim}(z_i^m, z_j^m)/\tau)}{\sum_{k=1}^{2N} \mathcal{I}_{[k\neq i]} \exp(\text{sim}(z_i^m, z_k^m)/\tau)}
\label{eq:contrast_loss_pro}
\end{equation} 
In this formulation, positive pairs consist of embeddings from the same instrument type across different mixtures processed with identical effects, while negative pairs include the same instrument with different effects processing, extending Equation~(\ref{eq:contrast_loss}) to the instrument level. Our final training objective combines both mixture-level and instrument-level contrastive losses:
\begin{equation}
\mathcal{L} = \lambda_{\text{mix}} \cdot \ell_{i,j}^{\text{mixture}} + \lambda_{\text{inst}} \cdot \ell_{i,j}^{\text{inst}}
\label{eq:final_loss}
\end{equation}
where $\lambda_{\text{mix}} + \lambda_{\text{inst}} = 1.0$. This dual-objective approach enables our model to simultaneously learn effects representations at both mixture and instrument-specific levels. We implement a curriculum learning strategy by starting with $\lambda_{\text{mix}} = 1.0$  and linearly introducing the instrument-wise objective over training step. This is necessary because instrument-wise gradients lack meaningful guidance until the model has established reasonably accurate mixture-level representations. Our final loss weighting uses $\lambda_{\text{mix}} = 0.8$ and $\lambda_{\text{inst}} = 0.2$, deliberately prioritizing mixture-level learning. This weighting reflects a critical insight: robust mixture-level effects representations form the foundation for successful instrument-specific extraction, as the extractor directly operates on these mixture-level embeddings to isolate individual components.

\subsection{Model Architecture and Training Details}

\noindent \textbf{Audio Processing Pipeline}. For effects normalization, we follow \cite{martinez2022automatic} but exclude reverb normalization. We implement audio processors using \texttt{dasp-pytorch} \cite{dasp} and \texttt{torchcomp} \cite{ycy2024diffapf}, incorporating seven effects types (equalizer, distortion, multiband-compressor, limiter, gain, stereo imager, delay). Reverberation uses convolution-based processing with in-house impulse responses, while using \texttt{pyloudnorm} for loudness normalization \cite{steinmetz2021pyloudnorm}.

\noindent \textbf{Model Components}. The encoder $\mathcal{E}$ is implemented using the PANN architecture \cite{kong2020panns}, following established practices in audio effects processing \cite{steinmetz2024st}. Input audio is preprocessed by computing log-melspectrograms with a window size of $2048$ samples and hop size of $512$ samples. We clip magnitude values between $-80$ and $40$ dB before scaling the spectrograms to the range $[-1, 1]$. Each input segment is $10$-seconds long to adequately capture audio effects characteristics. The \emph{extractor} mechanism employs a 3-layer MLP with hidden dimension $128$ and LeakyReLU activation (slope $0.1$). For instrument queries, we leverage the CLAP model \cite{wu2023large} to produce embeddings, applying a high dropout rate (0.75 to 0.95) during training to improve generalization and bridging the modality gap between text-audio encoders in CLAP, following the approach in \cite{saijo2025leveraging}. 

\noindent \textbf{Training Procedure} We train our model using the Adam optimizer \cite{kingma2014adam} with $\beta_1$ = 0.99 and $\beta_2$ = 0.9, employing a warm-up schedule \cite{goyal2017accurate} followed by cosine learning rate decay from a base rate of $1e-4$. We set contrastive temperature to $0.1$ and train on 2 NVIDIA H100 GPUs with a batch size of $192$. 
For our experiments, we employ the MoisesDB dataset \cite{pereira2023moisesdb}, which contains $240$ tracks with $11$ stems across $12$ genres, totaling over $14$ hours of multitrack data. It should be noted that tracks from the MoisesDB dataset \cite{pereira2023moisesdb} are ``wet,'' meaning they have already been processed with audio effects. To enhance model's robustness across varying mixture complexities, we randomly selected between one to four instruments for constructing each mixture during training.\footnote{\url{https://github.com/SonyResearch/Fx-Encoder_PlusPlus}}


\section{Evaluation Method}

\begin{table*}[h]
\centering
\footnotesize
\setlength{\tabcolsep}{2pt}
\begin{tabular}{l c | c c c c | c c c c | c c c c | c c c c | c c c}
\toprule
\multirow{3}{*}{Type} & \multirow{3}{*}{Model} & \multicolumn{19}{c}{MUSDB18 \cite{musdb18-hq}} \\
\cmidrule(lr){3-21}
& & \multicolumn{4}{c|}{Drums} & \multicolumn{4}{c|}{Bass} & \multicolumn{4}{c|}{Vocals} & \multicolumn{4}{c|}{Other} & \multicolumn{3}{c}{Mixture} \\
\cmidrule(lr){3-6} \cmidrule(lr){7-10} \cmidrule(lr){11-14} \cmidrule(lr){15-18} \cmidrule(lr){18-21}
& & R@1 & R@5 & R@10  & L$_{d}$ & R@1 & R@5 & R@10  & L$_{d}$ & R@1 & R@5 & R@10  & L$_{d}$ & R@1 & R@5 & R@10  & L$_{d}$ & R@1 & R@5 & R@10  \\
\midrule
\multirow{3}{*}{GP}  & CLAP & 1.5 & 3.9 & 5.9  & 1.38 & 1.4 & 3.7 & 6.0  & 1.39 & 0.7 & 3.0 & 4.7 & 1.67 & 0.3 & 1.8 & 3.4  & 1.40 & 1.6 & 5.2 & 8.7 \\
 & PANN & 0.8 & 2.2 & 3.7 & 1.75 & 0.8 & 2.9 & 4.6  & 1.59 & 0.3 & 1.8 & 3.1  & 1.75 & 0.3 & 0.9 & 2.3  & 1.54 & 1.2 & 3.8 & 6.5 \\
& VGGish & 2.0 & 5.7 & 8.5  & \textbf{1.13} & 1.8 & 4.7 & 6.7  & 1.28 & 1.1 & 3.7 & 5.8  & 1.65 & 0.6 & 2.6 & 4.3  & 1.39 & 1.4 & 5.1 & 7.8 \\
\midrule
\multirow{3}{*}{Fx}  & FX-Encoder & 12.1 & 25.9 & 35.1  & 1.33 & 7.3 & 17.8 & 24.9 & 1.39 & 6.1 & 16.0 & 22.8 & 1.92 & 5.8 & 15.4 & 22.5  & \textbf{1.37} & 12.3 & 26.6 & 34.5 \\
 & AFx-Rep & 20.8 & 34.8 & 40.5 & 1.15 & 13.1 & 24.9 & 31.8 & 1.18 & 11.5 & 23.1 & 29.3  & \textbf{1.42} & 10.9 & 21.9 & 27.1  & 1.40 & 16.7 & 30.2 & 36.9 \\
 & Fx-Encoder++ & \textbf{26.1} & \textbf{40.7} & \textbf{46.5}  & 1.16 & \textbf{22.1} & \textbf{35.9} & \textbf{42.0} & \textbf{1.13} & \textbf{14.2} & \textbf{25.4} & \textbf{31.9} & 1.58 & \textbf{17.4} & \textbf{30.2} & \textbf{37.1} & 1.38 & \textbf{19.4} & \textbf{33.5} & \textbf{40.5} \\
\midrule
\multirow{3}{*}{Type} & \multirow{3}{*}{Model} & \multicolumn{19}{c}{MedleyDB \cite{bittner2014medleydb}} \\
\cmidrule(lr){3-21}
& & \multicolumn{4}{c|}{Mandolin} & \multicolumn{4}{c|}{Alto Saxophone} & \multicolumn{4}{c|}{Horn} & \multicolumn{4}{c|}{Trumpet} & \multicolumn{3}{c}{Mixture} \\
\cmidrule(lr){3-6} \cmidrule(lr){7-10} \cmidrule(lr){11-14} \cmidrule(lr){15-18} \cmidrule(lr){18-21}
& & R@1 & R@5 & R@10  & L$_{d}$ & R@1 & R@5 & R@10  & L$_{d}$ & R@1 & R@5 & R@10  & L$_{d}$ & R@1 & R@5 & R@10  & L$_{d}$ & R@1 & R@5 & R@10  \\
\midrule
\multirow{3}{*}{GP}  & CLAP  & 1.4 & 3.5 & 5.8  & 1.24 & 1.5 & 4.2 & 6.9  & 1.43 & 1.9 & 4.5 & 6.1 & 3.1 & 1.36 & 1.2 & 4.5 & 1.42 & 0.8 & 3.4 & 5.4 \\
 & PANN & 0.6 & 2.5 & 4.9 & 1.24 & 0.8 & 2.4 & 4.3 & 1.53 & 1.4 & 3.4 & 4.9  & 1.33 & 0.9 & 2.2 & 3.3 & 1.42 & 0.7 & 2.4 & 4.9 \\
 & VGGish & 1.7 & 4.4 & 7.5  & 1.26 & 2.0 & 5.4 & 8.1 & 1.44 & 2.8 & 5.7 & 7.5 & 1.30 & 1.5 & 3.6 & 5.1  & 1.33 & 0.9 & 2.6 & 3.8 \\
\midrule
\multirow{3}{*}{Fx} & FX-Encoder & 2.3 & 6.3 & 10.1 & 1.27 & 2.2 & 6.5 & 10.0  & 1.38 & 3.4 & 8.5 & 11.9 & 1.33 & 1.7 & 4.8 & 6.8  & 1.36 & 0.8 & 3.0 & 5.0 \\
& AFx-Rep  & 15.8 & 25.2 & 30.8 & \textbf{1.15} & 12.7 & 21.3 & 26.1 & 1.37 & 12.2 & 19.9 & 24.4  & \textbf{1.17} & 8.0 & 14.3 & 17.7  & 1.40 & 4.2 & 10.1 & 13.6  \\
 & Fx-Encoder++ & \textbf{17.5} & \textbf{27.8} & \textbf{34.1} & 1.20 & \textbf{18.6} & \textbf{29.4} & \textbf{35.4} & \textbf{1.33} & \textbf{13.5} & \textbf{21.6} & \textbf{25.9}  & 1.24 & \textbf{9.1} & \textbf{16.1} & \textbf{19.8}  & \textbf{1.30} & \textbf{5.6} & \textbf{12.4} & \textbf{17.3}  \\
\bottomrule
\end{tabular}
\label{tab:isolated_inst_results}
\caption{Audio effects retrieval performance (R@K values in percentages, higher is better) on isolated instruments and mixtures. The $L_{d}$ column indicates performance in audio effects parameter matching tasks (lower is better). GP: general purpose audio representations. Fx: audio effects specific representations. }
\label{tab:isolated_inst_results}

\vspace{0.2cm}

\centering
\renewcommand{\arraystretch}{1.0}
\resizebox{1.0\linewidth}{!}
{
\setlength{\tabcolsep}{4pt}
\begin{tabular}{l c | c c c c | c c c c | c c c c}
\hline
\multirow{3}{*}{Type} & \multirow{3}{*}{Model} & \multicolumn{12}{c}{MUSDB18 \cite{musdb18-hq}} \\
\cmidrule{3-14}
& & \multicolumn{4}{c|}{Target Instrument (Oracle)} & \multicolumn{4}{c|}{\emph{USS}($m$) / \emph{MSS}($m$)} & \multicolumn{4}{c}{$\mathcal{E}(\mathbf{x, q_{audio})}$ / $\mathcal{E}(\mathbf{x, q_{text})}$} \\
\cmidrule(lr){3-6} \cmidrule(lr){7-10} \cmidrule(lr){11-14}
& & R@1 & R@5 & R@10 & mAP@10 & R@1 & R@5 & R@10 & mAP@10 & R@1 & R@5 & R@10 & mAP@10 \\
\hline
\multirow{3}{*}{GP} & CLAP & 1.0 & 3.1 & 5.0 & 2.0 & 0.3 / 0.3 & 1.2 / 1.5 & 2.4 / 2.8 & 0.8 / 0.8 &  - & - & - & - \\
& PANN & 0.6 & 2.0 & 3.4 & 1.2 & 0.3 / 0.3 & 1.1 / 1.4 & 2.3 / 2.4 & 0.7 / 0.8 &  - & - & - & - \\
& VGGish & 1.4 & 4.2 & 6.3 & 2.6 & 0.3 / 0.5 & 1.3 / 1.7 & 2.7 / 3.0 & 0.7 / 1.1 & - & - & - & - \\
\midrule
\multirow{3}{*}{Fx}  & FX-Encoder & 7.8 & 18.8 & 26.3 & 12.6 & 0.6 / 2.0 & 2.4 / 7.0 & 4.1 / 11.5 & 1.4 / 4.3 & - & - & - & - \\
 & AFx-Rep & 14.0 & 26.2 & 32.2 & 19.3 & \textbf{2.1} / 4.6 & 6.0 / 10.8 & 8.8 / 15.0 & 3.8 / 7.3 & - & - & - & - \\
 & Fx-Encoder++ & \textbf{19.9} & \textbf{33.0} & \textbf{39.4} & \textbf{25.6} & \textbf{2.1 / 7.1} & \textbf{6.4 / 15.4} & \textbf{9.9 / 20.5} & \textbf{4.0 / 10.8} & \textbf{3.0 / 3.0} & \textbf{8.1 / 8.1} & 11.7 / \textbf{12.2} & 5.2 / \textbf{5.3} \\
\hline
\multirow{3}{*}{Type} & \multirow{3}{*}{Model} & \multicolumn{12}{c}{MedleyDB \cite{bittner2014medleydb}} \\
\cmidrule{3-14}
& & \multicolumn{4}{c|}{Target Instrument (Oracle)} & \multicolumn{4}{c|}{\emph{USS}($m$)} & \multicolumn{4}{c}{$\mathcal{E}(\mathbf{x, q_{audio})}$ / $\mathcal{E}(\mathbf{x, q_{text})}$} \\
\cmidrule(lr){3-6} \cmidrule(lr){7-10} \cmidrule(lr){11-14}
& & R@1 & R@5 & R@10 & mAP@10 & R@1 & R@5 & R@10 & mAP@10 & R@1 & R@5 & R@10 & mAP@10 \\
\hline
\multirow{3}{*}{GP} & CLAP & 0.9 & 2.6 & 4.4 & 1.7 & 0.3 & 1.5 & 2.8 & 0.9 &  - & - & - & - \\
& PANN  & 2.0 & 4.8 & 7.0 & 3.2 & 0.3 & 1.3 & 2.5 & 0.8 & - & - & - & -  \\
& VGGish  & 1.5 & 3.8 & 5.8 & 2.6 & 0.4 & 1.4 & 2.5 & 0.8 & - & - & - & - \\
\midrule
\multirow{3}{*}{Fx} & FX-Encoder & 2.4 & 6.5 & 9.7 & 4.2 & 0.6 & 2.2 & 3.9 & 1.4 & - & - & - & - \\
 & AFx-Rep & 12.1 & 20.2 & 24.8 & 15.6 & \textbf{1.8} & \textbf{4.7} & \textbf{7.2} & \textbf{3.1} & - & - & - & - \\
 & Fx-Encoder++ & \textbf{14.7} & \textbf{23.7} & \textbf{28.8} & \textbf{18.7} & 1.7 & 4.6 & 7.0 & 3.0 & 1.6 / \textbf{1.9} & 5.0 / \textbf{5.7} & 7.6 / \textbf{8.7} & 3.1 / \textbf{3.6} \\
\hline
\end{tabular}
}
\caption{Audio effects retrieval performance (R@K values and mAP in percentages) on instrument-wise extraction.}

\label{tab:inst_wise_res}

\end{table*}

\subsection{Audio Effects Retrieval}
To evaluate our model, we conduct audio effects retrieval experiments using a controllable effects pipeline with MUSDB \cite{musdb18-hq} and MedleyDB \cite{bittner2014medleydb}.
Our evaluation builds upon AFx-Rep \cite{steinmetz2024st} and makes further improvements in three dimensions, including: (1) expanding small retrieval pools (previously limited to only 20 samples), (2) implementing effects normalization (addressing the issue of inherent effects in audio), and (3) extending performance metrics (beyond the previously reported accuracy). 

\noindent\textbf{Evaluation Framework.} Our retrieval task evaluates how accurately models identify correct effects configurations from 500 candidates when presented with processed audio. Given audio query, our goal is to retrieve the target audio with same effects configurations but different musical contents. The task uses cosine similarity in the embedding space between queries and candidate pools, following the evaluation methodology established in CLAP \cite{wu2023large}. We report Recall@K (R@K) in percentages, applying effects to evaluation datasets using the procedure described in Section \ref{section:objective_for_fx}. We construct separate evaluation sets from MUSDB \cite{musdb18-hq} (vocals, drums, bass, other) and from MedleyDB \cite{bittner2014medleydb} (mandolin, saxophone, horn, trumpet) to test generalization across instrument diversity. 
We further assess performance across effects complexity by varying the number of effects applied to each track from $1$ to $8$.

\noindent \textbf{Baselines.} We compare against general-purpose representations (VGGish \cite{hershey2017cnn}, PANN \cite{kong2020panns}, CLAP \cite{wu2023large}) and specialized effects representations (FX-Encoder \cite{koo2023music}, AFx-Rep \cite{steinmetz2024st}), excluding guitar-specific models \cite{chen2024towards, wright2025open} unsuitable for multi-instrument evaluation.

\begin{figure}[!th]
 \centerline{
 \includegraphics[width=0.9\columnwidth, keepaspectratio]{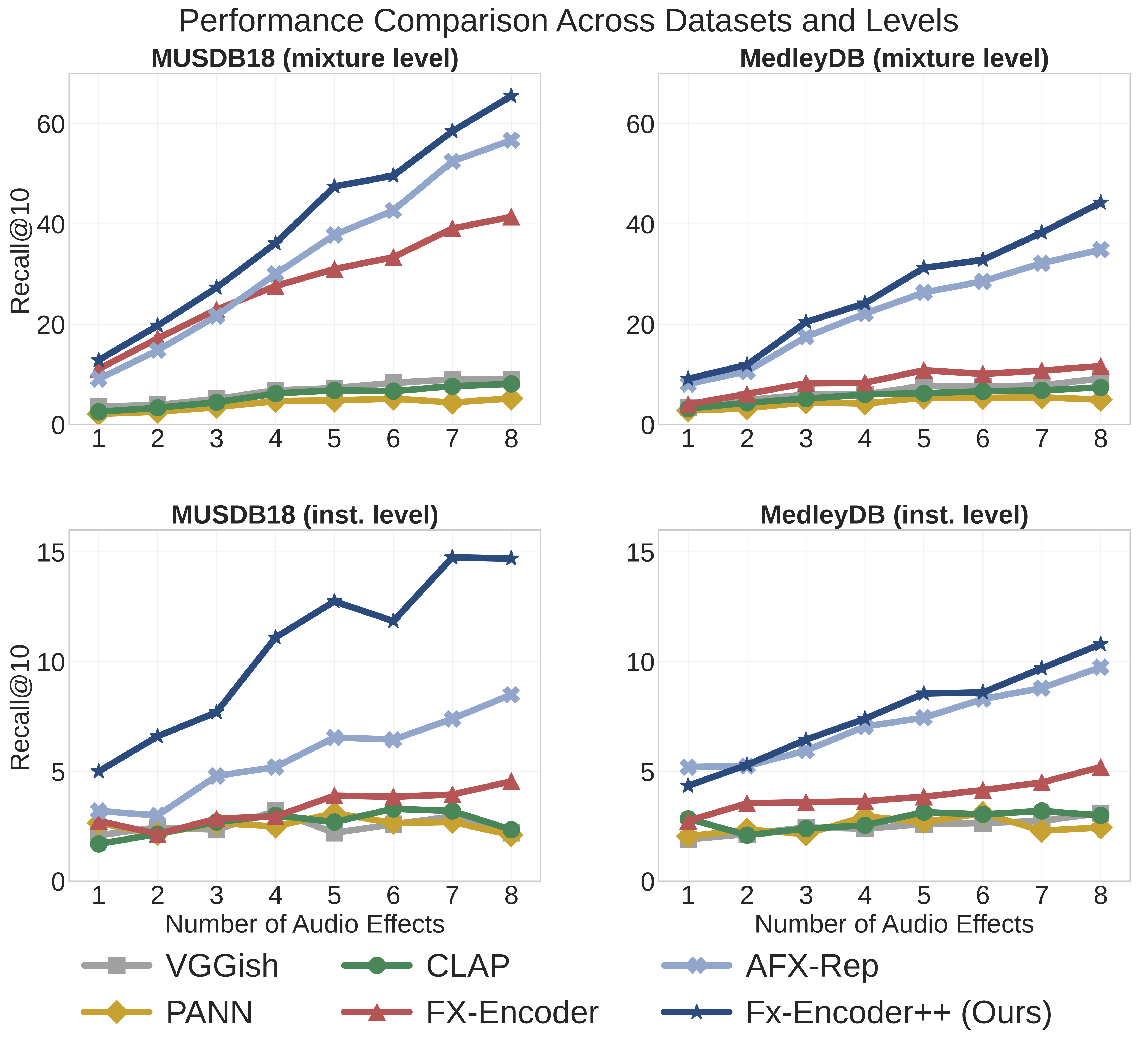}
 }
 \vspace{-5mm}
 \caption{Performance comparison across datasets, mixture and instrument-level, and effects complexity.}
 \label{fig:numfx_plot}
\end{figure}

\noindent \textbf{Retrieval Scenarios.} We evaluate using two complementary scenarios: (1) \emph{Mixture-level Retrieval}, testing effects identification in complete mixtures or isolated recordings \cite{koo2023music, steinmetz2024st}; and (2) \emph{Instrument-wise Retrieval}, evaluating our novel extraction capability with both ground-truth isolated tracks and directly from mixtures. For baseline models lacking instrument-specific capabilities, we implement a two-stage approach using universal sound separation \cite{kong2023universal} for arbitrary instrument types and Hybrid Demucs \cite{defossez2021hybrid} for MUSDB18's four standard tracks (vocals, bass, drums, other). In contrast, Fx-Encoder++ directly extracts instrument-specific embeddings without requiring separation. We support both audio and text queries through our CLAP encoder, using the prompt template: ``\emph{this is the sound of \{target instrument\}}'' \cite{saijo2025leveraging} for text queries.

\subsection{Matching of Audio Effects Parameters}
We evaluate downstream applications through effects parameters matching following \cite{steinmetz2024st}, extracting effects representations from reference audio 
to optimize differentiable effect chain parameters applied to dry input while minimizing a multi-resolution STFT Loss~\cite{steinmetz2020auraloss}. Our evaluation uses audio triplets (clean, reference, target) where input/target share identical content with different 
applied audio effects, while reference/target have different content but identical effects. We synthesize 100 samples per instrument from MUSDB \cite{musdb18-hq} and MedleyDB \cite{bittner2014medleydb} datasets, applying seven sequential effects (EQ, multiband compressor, stereo imager, gain, distortion, delay, limiter). We compare against the same baselines used in our retrieval experiments.

\vspace{-0.4cm}

\section{Results}

\subsection{Audio Effects Retrieval}

\textbf{Mixture-level Retrieval}. Table~\ref{tab:isolated_inst_results} shows Fx-Encoder++ substantially outperforms both general-purpose models 
and other effect-specific models 
on MUSDB dataset. Our model particularly excels with ``drums'' 
and ``bass'' 
. However, performance on ``vocals'' is comparatively lower than other instruments (drops approximately 10\%), likely due to the high timbral variation of vocals in each song (e.g. male vs. female singers). This pattern is also observed in the ``other'' and ``mixture'' categories, which similarly exhibit high timbral variety. All models show decreased performance on the MedleyDB dataset, suggesting that performance is limited by the frequency of instrument exposure during training. Despite this challenge, Fx-Encoder++ maintains its performance advantage across datasets.

\noindent \textbf{Instrument-wise Retrieval}. Table~\ref{tab:inst_wise_res} presents retrieval results across three protocols: (1) Target Instrument (Oracle): using ground-truth isolated tracks; (2) \emph{USS}($m$) / \emph{MSS}($m$): applying universal \cite{kong2023universal} or music source separation \cite{defossez2021hybrid}; and (3) $\mathcal{E}(\mathbf{x, q_{audio}})  / \mathcal{E}(\mathbf{x, q_{text}})$: our extractor applied directly to mixtures. For MUSDB dataset, our model substantially outperforms general-purpose models on target instruments. USS-based methods suffer significant performance drops (9.9\% R@10), while MSS performs better (20.5\% R@10) but is limited to only four predefined stems. Notably, even when using high-quality source separation, we observe a clear gap between the ``Target Instrument'' and ``\emph{MSS}($m$)'', (approximately 19\% lower), indicating that separation artifacts may distort the effects characteristics. 
Our extractor outperforms \textit{USS(m)} without requiring an external source separation model, with text queries perform slightly better than audio queries, probably because they provide a more generalized instrument space compared the $q_\text{audio}$, as the query audio has different content and Fx than the target.
For the MedleyDB dataset, performance degrades across all protocols, highlighting challenges in generalizing to instruments observed less frequently during training. 


\noindent \textbf{Number of Effects}. Figure~\ref{fig:numfx_plot} demonstrates our model's superior performance across varying effect counts. Performance improves with more effects for all models, indicating complex effect chains create more distinctive signatures than single effect, which may be confused with natural instrument timbres. For MUSDB dataset, the performance gap widens with increasing effects complexity. While similar trends appear for MedleyDB dataset, reduced performance indicates challenges in generalizing to novel timbral characteristics.

\vspace{-0.4cm}

\subsection{Matching of Audio Effects Parameters}
Table \ref{tab:isolated_inst_results} shows varied matching performance ($L_{d}$) across instruments. Unlike retrieval tasks, no model consistently outperforms across all scenarios. For MUSDB dataset, different models excel in different categories (VGGish on drums, Fx-Encoder++ on bass, AFx-Rep on vocals), with general-purpose VGGish surprisingly outperforming specialized models in several cases. For MedleyDB dataset, AFx-Rep and Fx-Encoder++ achieve the lowest reconstruction losses. These findings suggest that retrieval and effect parameter matching tasks benefit from different representational characteristics.

\vspace{-0.4cm}

\section{Conclusion}
We introduced \textbf{Fx-Encoder++}, the first model that extracts instrument-wise audio effects information from music mixtures. Our approach outperforms existing methods in audio effects retrieval at the mixture level while enabling instrument-wise effects embedding extraction. Limitations include reduced effect parameter matching performance, and difficulties with single effect understanding. Future work should address these challenges by improving retrieval-transformation bridging, enhancing single effect understanding for applications like VA modeling \cite{martinez2020deep, yeh2024hyper, steinmetz2021efficient}.

\section{Acknowledgements}
Yen-Tung thanks National Science and Technology Council for supporting his PhD study. 

\bibliography{ISMIRtemplate}

\end{document}